Non-Maxwellian electron distributions resulting from direct laser acceleration in near-critical plasmas

T. Toncian[1], C. Wang[1], E. McCary[1], A. Meadows[1], A.V. Arefiev[1], J. Blakeney[1], K. Serratto[1], D. Kuk[1], C. Chester[1], R. Roycroft[1], L. Gao[1], H. Fu[2], X.Q. Yan[2], J. Schreiber[3], I. Pomerantz[1], A. Bernstein[1], H. Quevedo[1], G. Dyer[1], T. Ditmire[1], B. M. Hegelich[1]

[1] Department of Physics, University of Texas, Austin, Texas 78712
[2] State Key Laboratory of Nuclear Physics and Technology, Peking University, Beijing 100871, China
[3] Fakultat fur Physik, Ludwig-Maximilians-University, Munich, Germany

Abstract

The irradiation of few nm thick targets by a finite-contrast high-intensity short-pulse laser results in a strong pre-expansion of these targets at the arrival time of the main pulse. The targets decompress to near and lower than critical densities plasmas extending over few micrometers, i.e. multiple wavelengths. The interaction of the main pulse with such a highly localized but inhomogeneous target leads to the generation of a short channel and further self-focusing of the laser beam. Experiments at the GHOST laser system at UT Austin using such targets measured non-Maxwellian, peaked electron distribution with large bunch charge and high electron density in the laser propagation direction. These results are reproduced in 2D PIC simulations using the EPOCH code, identifying Direct Laser Acceleration (DLA) [1] as the responsible mechanism. This is the first time that DLA has been observed to produce peaked spectra as opposed to broad, maxwellian spectra observed in earlier experiments [2]. This high-density electrons have potential applications as injector beams for a further wakefield acceleration stage as well as for pump-probe applications.

Introduction

The irradiation of solids with relativistic-intensity laser pulses leads to the generation of high-energy particles and radiation. In particularly the acceleration of electrons from solid targets has sparked interest due to the subsequent acceleration of ions or generation of higher harmonics of the driving laser wavelength. Thermal electron spectra are commonly generated by the irradiation of bulk targets [3,4], the specific details of the hereby-accelerated electron distribution have been a continuous topic of current research [5-7]. The density and number of electrons accelerated from a solid-density target is orders of magnitude higher than for under-dense plasmas from gas targets, though the electrons accelerated by the latter reach orders of magnitude higher energies and can exhibit very narrow, quasi mono-energetic spectral distributions. Recently, quasi mono-energetic spectra have been reported from the irradiation of a nanometer scale solid foil targets with a high-intensity and high-contrast laser pulse [8]. The results have been associated with the onset of the radiation pressure accelerations regime, and the produced dense electron bunched proposed to be utilized as relativistic moving mirrors upshifting a reflected optical laser pulse into the X-Ray range [5,9-14].
Here we demonstrate experimentally the generation of quasi mono-energetic electron spectra employing a few TW, moderated contrast laser system irradiating ultra thin, nanometer scale targets.

**Experimental Setup**

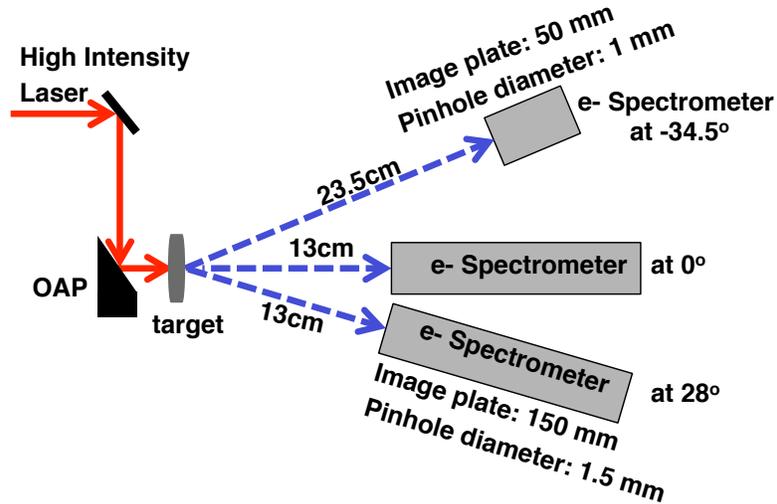

**Figure 1 Experimental setup. A linearly polarized laser pulse is focused by an f/2.8 off-axis-parabola (OAP) mirror on targets at normal incidence. Electrons emitted from the back of the target are detected with electron spectrometers located at three different angles.**

The experiment was performed using the Glass Hybrid OPCPA Scaled Test-bed (GHOST) laser at the Center for High Energy Density Science at the University of Texas at Austin. GHOST is based on an optical parametric chirped pulse amplification (OPCPA) and mixed glass amplifier chain. Typical pulse parameters for Ghost are pulse energy ~1J on target, pulse duration ~120fs and focal spot sizes of Ghost are ~5.5 μm FWHM diameter, resulting in on target intensity of ~$10^{19}$ Wcm$^{-2}$, equivalent to a normalized laser vector potential of $a_L$ ~4, where $a_L = I_L[\text{W/cm}^2]\lambda^2[\mu m^2]/(1.37 \times 10^{18})$, $I_L$ is the laser intensity, and $\lambda$ =1055 nm is the laser center wavelength.

The experimental set up is shown in Figure 1. The laser was focused onto Polyethylene foils with 500 nm thickness and Diamond Like Carbon (DLC) foils with thickness of 5nm and 20 nm. At solid densities, DLCs reach electron densities of $n_e \approx 10^{24}$ cm$^{-3}$ after full ionization. They where fabricated using a specialized cathode arc discharge [15].

Magnetic electron spectrometers covering three different angles were used to measure the electron distribution in the range of 100 keV to 15 MeV. The magnetic spectrometer was used in conjunction with image plates as detectors and the response of the system was absolutely calibrated.

**Experimental Results**

Figure 2 a-c shows the electron spectra generated from 500 nm plastic, 20 nm and 5nm DLC targets measured at 0 deg, 28 deg and -34.5 deg with respect to the laser propagation direction. Figure 2d combines the spectra at the laser incident direction from the above three different targets. Exponentially decaying distributions are measured for all angles from the 500 nm plastic film (see Figure 2a). However, there is an obvious deviation from an exponentially decaying distribution for the 20 nm thick DLC target (see Figure 2b) in the target normal direction, with a peaked electron distribution between 4.7 MeV and 7.8 MeV. The cutoff energy of the electron distribution is ~10 MeV with an energy spread of ~8 MeV. Additionally, we obtained a quasi-mono energetic electron peaked distribution from a 5 nm DLC film target with an energy peak at 3.8 MeV (see Figure 2d). This electron peak has an energy spread, which is about three times narrower as compared to the peaked electron distribution from the 20 nm thin DLC film target, whereas the peak electron number per MeV is comparable.

The angular distributions of the electron spectra from DLC targets exhibit reduced cut-off energies with increasing angles, while for the 500 nm target the cut off energy remains constant for all measured angles.

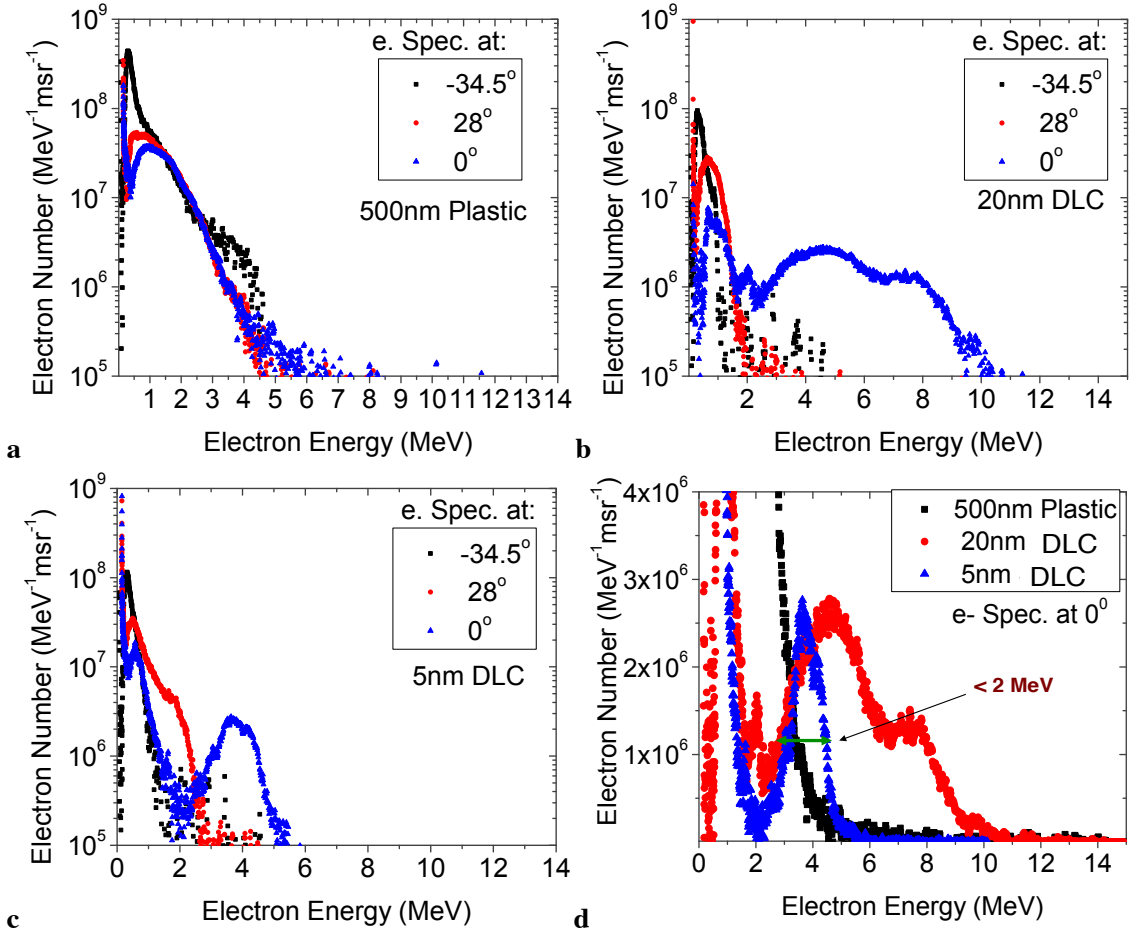

**Figure 2 Laser-driven electron spectra as measured from ultrathin film targets. a) Electron spectra from a 500 nm plastic film; b) Electron spectra from a 20 nm DLC film; c) Electron spectra from a 5 nm DLC film; For each type of target, the electron spectrometers (e- Spec. s) are located at -34.5° (black square), 0° (blue triangle), and 28° (red circle) with respect to the target normal; d) Electron spectra at target normal direction from the above three different targets.**

**Simulation Results**

For further interpretation of the experimental data we need to consider the realistic conditions of the laser plasma interaction, in particular the effect of the finite laser contrast. With a 3$^{rd}$ order scanning autocorrelator we have measured the parametric fluorescence pedestal generated due to the OPA laser architecture at constant relative intensity $I/I_{max}=5\times10^{-7}$ preceding the main laser pulse as long as 8 ns. With a sustained intensity of $2\times10^{13}$ W/cm$^2$, a severe decompression of the target should be expected. To estimate the target expansion we have utilized the HYADES code to simulate qualitatively the hydrodynamic heating and subsequent expansion of the target. Figure 3 shows the calculated ion density expansion as a function of time for a Polyethylene target. While the bulk of a 500 nm thick target remains at solid density with a pre-plasma expanding at the front side of the target (Figure 3a), an initial 20 nm thick target decompresses in a Gaussian like bell shape density distribution (Figure 3b). We expect for the calculation employing a Polyethylene target to overestimate the expansion due to a different EOS compared to a DLC target (EOS is not available in the code).

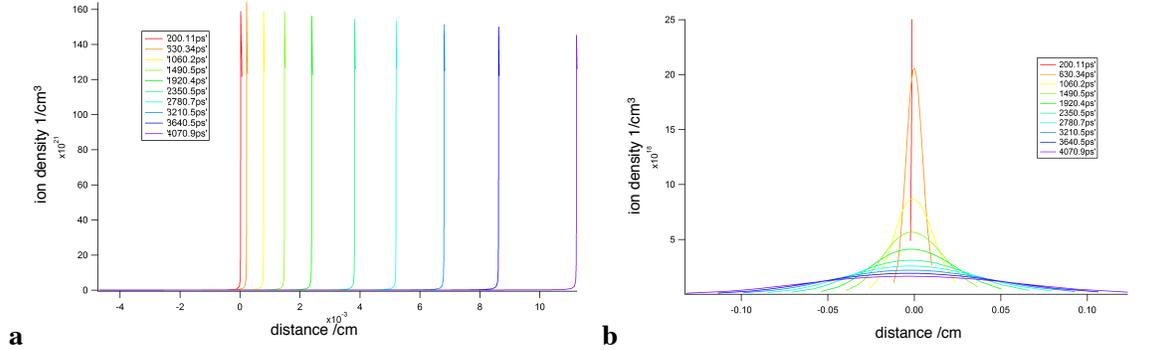

**Figure 3: Hyades hydrodynamic expansion simulation for an initial target thickness of a) 500 nm and b) 20 nm Polyethylene.**

We have conducted several two dimensional particle-in-cell simulations employing the EPOCH 4.5 framework. The peak laser intensity used in the simulations was $5\times10^{19}$ W/cm$^2$ corresponding to a $a_0 = 6$, in a linear polarized Gaussian laser pulse with a temporal FWHM of 100 fs and 3 $\mu$m focal spot size at normal incidence to the target. To take the target decompression into account we have repeated simulations with various target thicknesses while keeping the total areal mass constant. For a target with an initial thickness $L$ and initial density $n_0 = 10^{24}$/cm$^3$ expanding in a Gaussian like distribution (a width σ) the maximal density scales: $n_0 L = n_{max}\sigma\sqrt{2\pi}$. The electron distribution in the simulation box was evaluated 120 fs after the arrival of the peak intensity when the laser-plasma interaction is effectively over for three experimental measured angles of 0, 28 and -35 deg with respect to the laser propagation axis. Figure 4 shows the electron spectral distribution for the three targets employed in the experiment with an initial thickness of 500 nm, 20 nm and 5 nm and target widths σ of 1 $\mu$m, 2.5 $\mu$m and 5 $\mu$m. A qualitative agreement of the shape of the electron spectral distributions can be found for the 500 nm thick target for all simulated expansions, while for the 20 nm and 5 nm thin targets the σ = 5 $\mu$m expansion fits the observed distributions shape best. For the thin targets we observe a transition as the maximal target density is decreased. The electron distribution changes from an isotropic and exponentially decaying to one that is peaked in the forward direction and has a pronounced peak in the spectrum. This transition is observed as the plasma density becomes under-dense and the laser is able to propagate through the plasma.

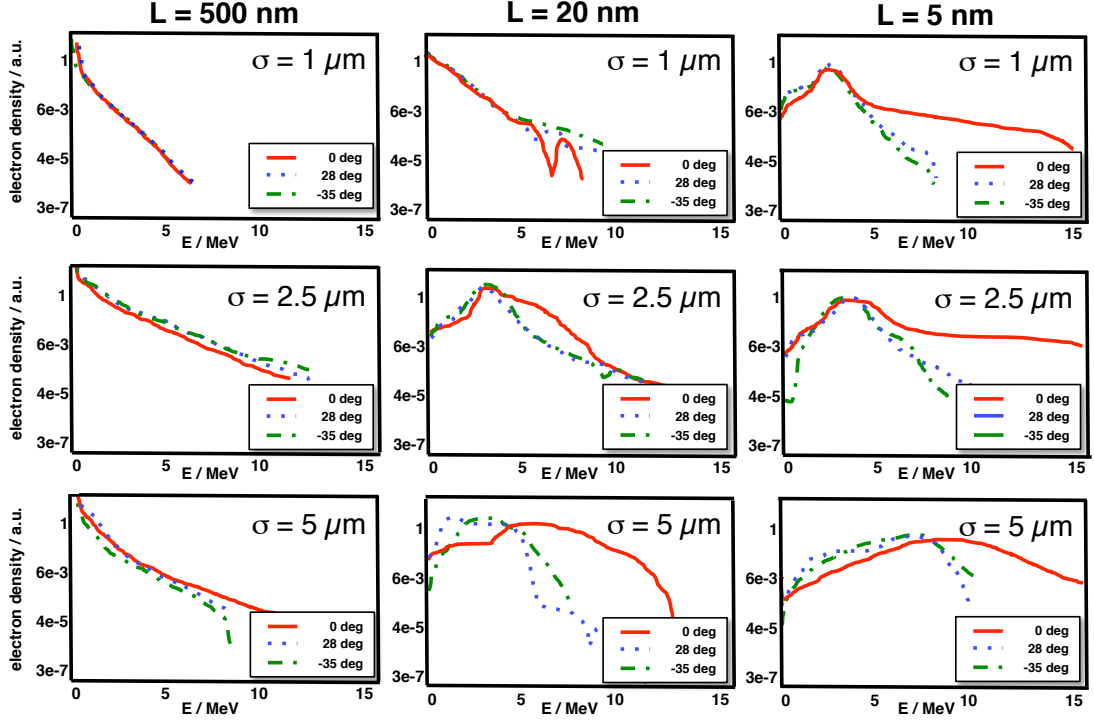

**Figure 4 Simulated electron spectral distributions for initially 500 nm, 20 nm and 5 nm (columns from left to right) thin targets pre-expanded to 1 μm, 2.5 μm and 5 μm. Spatially and spectrally peaked distributions are observed at near-critical electron densities.**

Non-thermal electron distributions are common results in the underdense plasma laser interactions [2,4,16-20]. Wakefield acceleration in the broken wave regime also know as bubble regime will accelerate quasi mono-energetic electron spectral distributions [21-25], but the short length of the plasma employed in our setup does not allow the onset of these electron acceleration mechanisms.

To further examine the details of the electron acceleration mechanism observed in our simulations, we have tracked a subset of electrons. Figure 5a shows a snapshot of the mean kinetic energy of the electrons in each simulation cell and highlights an electron bunch. Figure 5b illustrates the trajectories of electrons reaching an energy corresponding to a relativistic $\gamma = 30$ in the highlighted bunch (the trajectories are color coded with respect to $\gamma$). We observe that, although the electrons are captured into the laser pulse on the front side of the target from various locations, the highest $\gamma$ is reached while exiting at the rear side of the target. We have picked one of the electrons and reconstructed its acceleration history, with the resulting components of the momentum shown in Figure 5b as functions of time. The electron predominantly gains energy from the transverse electric field, which is the field of the laser. The Lorentz force then bends the electron trajectories towards laser direction, leading to longitudinal motion of the most energetic electrons. Figure 5b also shows the time-averaged longitudinal electric field. It decelerates the energetic electrons as they leave the target. After an injection phase, the electron moves in phase with the laser pulse gaining energy by the direct laser acceleration mechanism. The small size of the target and peaked density at the center limiting the injection volume are possibly responsible for in a peaked electron distribution.

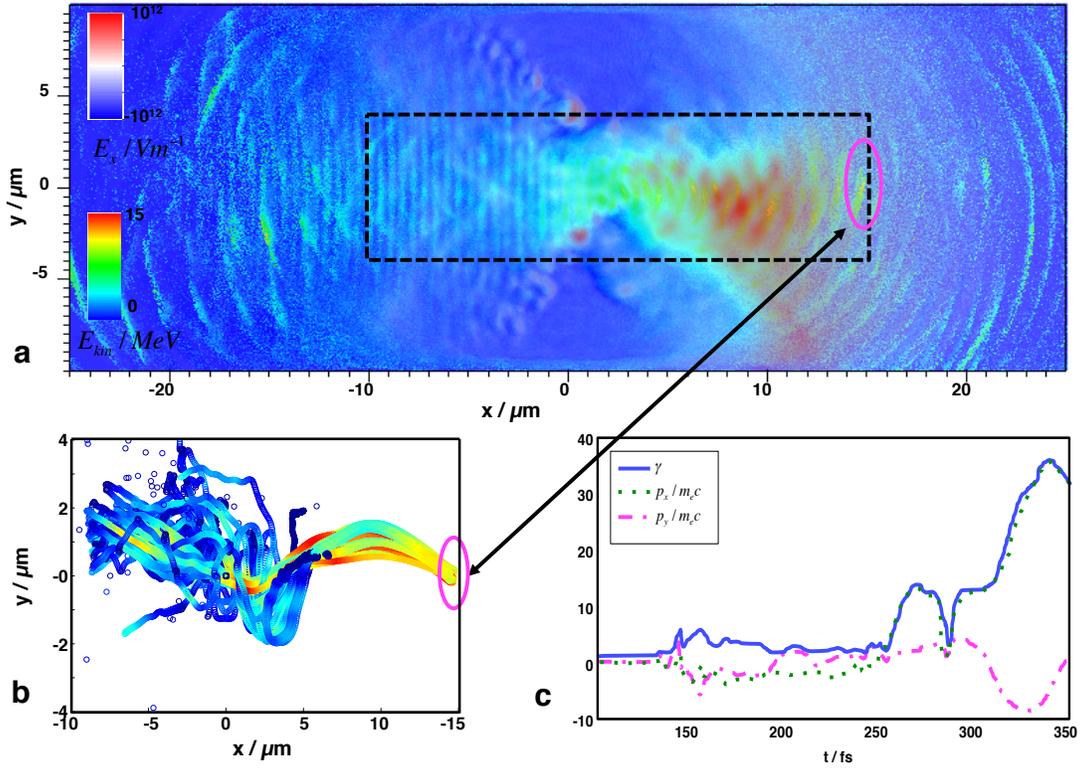

**Figure 5** Shows a) snapshot of the average kinetic energy of the electrons in each simulation cell and cycle average electric field in x-direction. b) shows trajectories of the electrons forming the bunch highlighted in a. c) shows the partition of transverse and longitudinal momentum in respect to the laser direction and resulting relativistic gamma factor.

**Conclusions and Outlook**

We have demonstrated experimentally that dense non-Maxwellian electron distributions can be generated by the irradiation of ultra thin targets with a finite contrast laser pulse. The laser energy preceding the main intensity spike results in a strong decompression of the target at the arrival time of the main pulse. For initial target thicknesses that have expanded to electron densities lower than the (relativistic) critical density the laser pulse will propagate through and interact with a finite plasma region. The limitation of the target length to several ten micrometers leads both to a finite acceleration length and a localization of the electron injection in to the laser wave. The hereby-generated electron distributions deviate from a thermal distribution exhibiting a distinct peaked spectral and spatial distribution. A further acceleration stage of these electrons can be envisioned in the following scenario, similar to the proposal by Zhang et al. [26], but with the added benefit of non-thermal, peaked electrons. A gas cell enclosed by a thin DLC window could be irradiated by a medium contrast short-pulse laser. High-density electron bunches generated by the laser interaction with the DLC window reach on a short spatial scale high $\gamma$ while moving in phase with the laser. Subsequent seeding into laser wakefield acceleration stage in the low density plasma contained by the gas cell will further accelerate the electrons. The feasibility of injection into a wakefield acceleration of the electrons in respect to beam loading and dephasing should be the scope of further investigations.